\documentclass[fleqn,usenatbib,twocolumn]{mnras}

\usepackage{newtxtext,newtxmath}

\usepackage[T1]{fontenc}

\DeclareRobustCommand{\VAN}[3]{#2}
\let\VANthebibliography\thebibliography
\def\thebibliography{\DeclareRobustCommand{\VAN}[3]{##3}\VANthebibliography}

\usepackage{subfig,graphicx}	
\usepackage{amsmath}	
\usepackage{booktabs}
\usepackage{float}
\usepackage{placeins}
\usepackage{outlines}
\usepackage{enumitem}
\usepackage[english]{babel}
\usepackage{blindtext}
\usepackage{tabularx}

\usepackage{xcolor, soul}

\title[USSRH in SPT-CLJ2337$-$5942]{Discovery of a $z 
\sim 0.8$ Ultra Steep Spectrum Radio Halo in the MeerKAT-South Pole Telescope Survey}

\author[Magolego et al.]{
Isaac S.~Magolego$^{1}$\thanks{E-mail: isaacike07@gmail.com}, 
Roger P.~Deane$^{1,2}$, 
Kshitij Thorat$^{2}$, 
Ian Heywood$^{3,4,5}$, 
William Rasakanya$^{1}$,
\newauthor
\hspace*{0.2em}Manuel Aravena$^{6,7}$, Lindsey E.~Bleem$^{8,9,10}$, Maria G.~Campitiello$^{8}$, Kedar A.~Phadke$^{11,12,13}$, Justin Spilker$^{14}$, 
\newauthor
\hspace*{0.2em}Joaquin D.~Vieira$^{11,12}$, Dazhi Zhou$^{15}$, Bradford A.~Benson$^{9,10,16}$, Scott Chapman$^{17,18,19}$, Ana Posses$^{14}$, 
\newauthor
\hspace*{0.2em}Tim Schrabback$^{20}$, Anthony Stark$^{21}$, David Vizgan$^{11}$\\\\
$^{1}$Wits Centre for Astrophysics, School of Physics, University of the Witwatersrand, Private Bag 3, 2050, Johannesburg, South Africa\\
$^{2}$Department of Physics, University of Pretoria, Hatfield, Pretoria, 0028, South Africa\\
$^{3}$Astrophysics, Department of Physics, University of Oxford, Keble Road, Oxford, OX1 3RH, UK\\
$^{4}$Department of Physics and Electronics, Rhodes University, PO Box 94, Makhanda 6140, South Africa\\
$^{5}$South African Radio Astronomy Observatory, 2 Fir Street, Observatory 7925, South Africa\\
$^{6}$Instituto de Estudios Astrof\'{\i}cos, Facultad de Ingenier\'{\i}a y Ciencias, Universidad Diego Portales, Av. Ej\'ercito 441, Santiago, Chile\\
$^{7}$Millenium Nucleus for Galaxies (MINGAL)\\
$^{8}$High-Energy Physics Division, Argonne National Laboratory, 9700 South Cass Avenue., Lemont, IL, 60439, USA\\
$^{9}$Kavli Institute for Cosmological Physics, University of Chicago, 5640 South Ellis Avenue, Chicago, IL, 60637, USA\\
$^{10}$Department of Astronomy and Astrophysics, University of Chicago, 5640 South Ellis Avenue, Chicago, IL, 60637, USA\\
$^{11}$Department of Astronomy, University of Illinois, 1002 West Green Street, Urbana, IL 61801, USA\\
$^{12}$Center for AstroPhysical Surveys, National Center for Supercomputing Applications, Urbana, IL, 61801, USA\\
$^{13}$NSF-Simons AI Institute for the Sky (SkAI), 172 E. Chestnut St., Chicago, IL 60611, USA\\
$^{14}$Department of Physics and Astronomy and George P. and Cynthia Woods Mitchell Institute for Fundamental Physics and Astronomy,\\
Texas A\&M University, 4242 TAMU, College Station, TX 77843-4242, USA\\
$^{15}$Department of Physics and Astronomy, University of British Columbia, 6225 Agricultural Road, Vancouver V6T 1Z1, Canada\\
$^{16}$Fermi National Accelerator Laboratory, MS209, P.O. Box 500, Batavia, IL 60510-0500\\
$^{17}$Department of Physics and Astronomy, University of British Columbia, 6225 Agricultural Road, Vancouver V6T 1Z1, Canada\\
$^{18}$National Research Council, Herzberg Astronomy and Astrophysics, 5071 West Saanich Road, Victoria V9E 2E7, Canada\\
$^{19}$Department of Physics and Atmospheric Science, Dalhousie University, 6310 Coburg Road, Halifax B3H 4R2, Canada\\
$^{20}$Universit\"at Innsbruck, Institut f\"ur Astro- und Teilchenphysik, Technikerstr. 25/8, 6020 Innsbruck, Austria\\
$^{21}$Harvard-Smithsonian Center for Astrophysics, 60 Garden Street, Cambridge, MA 02138, USA\\
}

\date{Accepted XXX. Received YYY; in original form ZZZ}

\pubyear{2015}

\begin{document}
\label{firstpage}
\pagerange{\pageref{firstpage}--\pageref{lastpage}}
\maketitle

\begin{abstract}
Radio halos are diffuse synchrotron sources that trace the turbulent intracluster medium (ICM) of galaxy clusters. However, their origin remains unknown. Two main formation models have been proposed: the hadronic model, in which relativistic electrons are continuously injected by cosmic-ray protons; and the leptonic turbulent re-acceleration model, where cluster mergers re-energise electrons in situ. A key discriminant between the two models would be the existence of ultra-steep spectrum radio halos (USSRHs), which can only be produced through turbulent re-acceleration. Here we report the discovery of an USSRH in the galaxy cluster SPT-CLJ2337$-$5942 at redshift $z = 0.78$ in the MeerKAT-South Pole Telescope 100~deg$^2$ survey. This discovery is noteworthy for two primary reasons: it is the highest redshift USSRH system to date; and the close correspondence of the radio emission with the thermal ICM as traced by {\sl Chandra} X-ray observations, further supporting the leptonic re-acceleration model. The halo is under-luminous for its mass, consistent with a minor merger origin, which produces steep-spectrum, lower luminosity halos. This result demonstrates the power of wide-field, high-fidelity, low-frequency ($\lesssim1$~GHz) surveys like the MeerKAT-SPT 100 deg$^2$ programme to probe the origin and evolution of radio halos over cosmic time, ahead of the Square Kilometre Array.
\end{abstract}




\section{Introduction} \label{sec:intro}

Galaxy clusters are the largest gravitationally bound structures in the universe. Their halo masses are of order 10$^{14-15}$ M$_\odot$ and are primary nodes in the cosmic web, a network of galaxy-filled filaments stretching across tens to hundreds of megaparsecs \citep[Mpc;][]{Arag_n_Calvo_2010,Cautun_2014,Malavasi_2020}. Clusters are important cosmic laboratories for both astrophysics and cosmology. They are predominately comprised of dark matter, which accounts for about 70-80$\%$ of their mass, while 15-20$\%$ is hot (10$^{6-7}$~K) ionized gas known as the intracluster medium \citep[ICM;][]{Blumenthal_1984,White_1995,Jones_1999,Vikhlinin_2006,Arag_n_Calvo_2010}. The remaining fraction lies within stars and galaxies. Occasionally, clusters collide and merge in powerful events, releasing huge amounts of energy ($\sim$10$^{64}$ erg) over timescales of order several Gyr. This energy creates shocks and turbulence that heat the ICM, which can leave signatures of the cluster merger activity \citep{Markevitch_2007}. Based on their state, clusters can be classified as either "relaxed" (undisturbed), or "merging" (disturbed), which are disturbed by these energetic interactions \citep{Markevitch_2007,Mann_2012,Yuan_2020}. One complementary approach to studying a cluster's dynamics is by detecting diffuse radio synchrotron emission, generated in the presence of large-scale weak magnetic fields ($\sim1~\mu$G), that extend across cluster environments. This emission, typically classified as radio halos and relics, is often associated with merger activity within the cluster \citep[e.g.,][]{Feretti_2012,Brunetti_2014,Weeren_2019}.

Radio relics are elongated structures typically found at the outskirts of galaxy clusters. They exhibit polarized emission and their elongation (major axis) is oriented perpendicular to the line connecting the relic to the cluster centre. These relics, typically found in disturbed systems \citep[][and references therein]{Weeren_2019}, challenge our understanding as they have physical sizes (Mpc-scale) significantly larger than expected considering the radiative lifespan ($t \sim 10-100$~Myr) of radiating particles emitting at radio wavelengths \citep{Bagchi_2002}. These characteristics suggest a cluster merger origin, where particles are re-energized through resultant shocks, one proposed model for which is diffusive shock acceleration \citep[DSA;][]{Ensslin_1997,Roettiger_1999}. However, some radio relics pose a challenge to our understanding of the DSA model, particularly (a) inferred low Mach numbers are inconsistent with existing re-acceleration/DSA models \citep{Macario_2011}; (b) clusters featuring strong X-ray shocks show no relic emission \citep{Russell_2011}; and (c) the alignment of magnetic fields within specific relics remains inconsistent with theoretical predictions in certain cases \citep{Weeren_2010}.

In contrast, radio halos are diffuse synchrotron emission components that span approximately a megaparsec in size, primarily co-located at the centre of the galaxy clusters. Their surface brightness typically measures in the range of a few $\mu$Jy/arcsec$^{2}$ at 0.1 - 1 GHz, making them orders of magnitude fainter compared to their relic counterparts. These halos have been identified in $\sim$ 50 high-mass clusters \citep[e.g.,][and references therein]{Weeren_2019}, mostly at low to intermediate redshift ($z$ $\lesssim$ 0.4), with only a few identified beyond $z\approx 0.5$ \citep[see][for details and additional references]{Yuan_2015,Cuciti_2021,Pasini_2024,phuravhathu2025}. Similar to radio relics, the observed scales of radio halos suggest a need for particle re-acceleration \citep{Cassano_2008}. Two leading theoretical models are the hadronic or secondary-electron model, where relativistic electrons are generated from proton-proton collisions within the intracluster medium \citep{Dennison_1980, Blasi_1999}; and the primary electron or turbulent re-acceleration model. The latter asserts that an existing population of cosmic ray electrons undergoes re-acceleration due to turbulence generated by cluster mergers \citep{Ensslin_1997, Brunetti_2004}. While the possibility of combining the two models has not been excluded \citep{Brunetti_2011}, the turbulent re-acceleration model is currently favoured \citep{Brunetti_2000,Cassano_2010,Miniati_2015,Bruno_2021,Duchesne2021}. This preference arises from several key observational tensions with the hadronic model. First, the expected level of gamma-ray emission associated with neutral pion decay (a byproduct of hadronic collisions) has not been detected by instruments such as Fermi-LAT, placing stringent upper limits on the cosmic-ray proton energy density in clusters \citep{Ackermann_2010, Brunetti_2012, Zandanel_2014}. Second, the hadronic model predicts a smoother and more centrally peaked radio morphology than what is typically observed for radio halos \citep{Donnert_2010}. Third,  the hadronic model has difficulties in reproducing the observed bimodality in the radio-X-ray luminosity relation and fails to account for the existence of ultra-steep spectrum halos ($\alpha \sim 1.5$-1.9)\footnote{$S_{v}$ $\propto$ $v^{-\alpha}$, where $S_{v}$ represents the source flux density at frequency $v$, and $\alpha$ is the spectral index used in this study.} detected predominantly at low radio frequencies \citep{Cassano_2006, Brunetti_2008}. These ultra-steep spectrum radio halos (USSRHs), are therefore a crucial class of objects for discerning between different models of the radio halo formation mechanism. USSRH detections are limited to the relatively low redshift universe at present, and often have low signal-to-noise (SNR) ratios and limited fidelity. 

Observations at high redshift are crucial for testing theoretical models, studying their evolution, and exploring their connection to the intracluster medium (ICM). In particular, detections of USSRHs at higher redshift are important because turbulent re-acceleration models predict that halos become rarer and fainter at earlier cosmic epochs (e.g. \citealt{Cassano_2010b}; \citealt{Brunetti_2014}), when clusters are dynamically younger and inverse-Compton (IC) losses against the cosmic microwave background
(CMB) are stronger (e.g. \citealt{Brunetti_2007}; \citealt{Cassano_2012}, \citealt{Gennaro2021}). Finding USSRHs at high-z therefore provides unique constraints on the efficiency of particle re-acceleration, the role of cluster mergers, and the evolution of cluster magnetic fields over cosmic time.

In this Letter, we report the discovery of the most distant ultra-steep-spectrum radio halo (USSRH) known to date, detected in the MeerKAT-South Pole Telescope survey (UHF band) at the position of the galaxy cluster SPT-CLJ2337$-$5942. This SZ-selected cluster was first discovered by \citet{Vanderlinde_2010} and lies within the South Pole Telescope (SPT) 100 square degree (deg$^2$) deep field \citep{Huang_2020}. It has a redshift of $z = 0.78$ and a mass of $M_{500} = 8.32 \times 10^{14}$ M${\odot}$ \citep{Huang_2020}, where $M_{500}$ is the mass enclosed within a sphere of radius $R_{500}$, defined as the radius within which the average density is 500 times the critical density of the Universe at the cluster’s redshift. This Letter is structured as follows: Section~\ref{sec2} briefly outlines the observations and data processing, Section \ref{sec3} describes the discrete source subtraction technique and low-resolution high brightness temperature ($T_B$) imaging, Section \ref{sec4} discusses the results, and Section \ref{sec5} presents our conclusions. We adopt a $\Lambda$CDM flat cosmology with $H_{0} = 70~\mathrm{km}~\mathrm{s}^{-1}~\mathrm{Mpc}^{-1}$, $\Omega_{m} = 0.3$, and $\Omega_{\Lambda} = 0.7$. At the redshift of SPT-CLJ2337$-$5942 ($z = 0.78$), $1''$ corresponds to 7.42 kpc.
\begin{table*}
\caption{The properties of SPT-CLJ2337$-$5942. The X-ray properties are adopted from \citet{Andersson_2011}.}
\small 
\centering
\begin{tabular}{ccccccc}  
\toprule\toprule
\textbf{Cluster} & \textbf{RA (J2000)} & \textbf{Dec (J2000)} & \textbf{$z$} & \textbf{$M_{500c}$} & \textbf{$L_{x, 500}$} & \textbf{$P_{816}$}\\
\midrule
& \text{(hms)} & \text{(dms)} & & \text{(10$^{14}$~M$_{\odot}$)} & \text{(10$^{44}$ erg s$^{-1}$)} & \text{(10$^{24}$~W\,Hz$^{-1}$)}\\
\midrule
SPT-CLJ2337$-$5942 & 23 37 24.2 & $-$59 42 17 & 0.78 & 8.32 $\pm$ 0.82 & 8.9 $\pm$ 0.5  & 2.5 $\pm$ 0.2\\
\bottomrule
\label{properties}
\end{tabular}
\begin{flushleft}
\textit{Note.} $P_{816}$ is the radio power of the USSRH at 816 MHz, the central frequency of the MeerKAT UHF-band.
\end{flushleft}
\end{table*}

\section{Observations and Data Processing}
\label{sec2}
\subsection{MeerKAT observations}
The radio observations were performed with the MeerKAT telescope, an array composed of 64 $\times$ 13.5 m antennas in the Northern Cape, South Africa. Approximately 70 per cent of the collecting area (48 antennas) is located in the central $\sim1$-km diameter core region, with the remainder spread out to longer baselines of up to 8~km \citep{Jonas_2016}. This provides excellent brightness temperature sensitivity, while still retaining the ability to calibrate on and subtract out compact sources in the visibility domain. MeerKAT currently has three frequency bands (UHF: 0.58 - 1.09 GHz, L-band: 0.9 - 1.7 GHz, S-band: 1.75 - 3.5 GHz), with plans to extend to a fourth frequency band from 8.4 to 15.4 GHz.\\\\
The 100 square degree (deg$^2$) MeerKAT UHF survey (“The MeerKAT-South Pole Telescope Survey,” Proposal ID:SCI-20220822-JV-02 and SCI-20230907-JV-01) is centered on \(\alpha = 23h30m\), \(\delta = -55^\circ\). The survey is comprised of 78 pointings, each with a primary beam FWHM of $\sim$1.6~deg at 816~MHz. Each pointing was observed for $\sim$ 1 hour, achieving a per-pointing depth of 10~$\mu$Jy beam$^{-1}$ and brightness temperature sensitivity of $\sim$ 258 mK, for Briggs-weighted (\textsc{Robust} = -0.5) images. The combined mosaic has a sensitivity of 14~$\mu$Jy beam$^{-1}$ when smoothed to a common resolution of 10.2 arcsec. The total survey observation time was 116~hr and used an 8-second integration time in the 32k correlator mode setup, resulting in 32,768 channels, each 16.602 kHz wide. The combined data volume is approximately $\sim$250 TB. Out of 64 MeerKAT antennas, 60 were typically available for each pointing. The sources J0408-6545 (primary calibrator) and J2329-4730 (secondary calibrator) were used for absolute flux, bandpass, phase and complex gain calibration.

\subsection{Calibration and Imaging}

We processed the MeerKAT data with the \textsc{Oxkat}\footnote{\url{https://github.com/IanHeywood/oxkat}} pipeline \citep{Heywood_2020} in combination with modified scripts of procedures of our own. \textsc{Oxkat} is a set of \textsc{Python}-based scripts that semi-automatically process Stokes I MeerKAT data, including direction-dependent calibration, which is essential for our large field-of-view, sensitivity, and $\lesssim$1~GHz observing frequency. Outlined below is a brief description of the steps we performed using the \textsc{Oxkat} pipeline.\\\\
The MeerKAT UHF-band data were processed using a combination of standard and advanced calibration techniques. First-generation calibration (1GC) was performed using the \textsc{Oxkat} pipeline in \textsc{Casa}\footnote{\url{https://casa.nrao.edu/}} \citep{McMullin2007}, reducing the number of channels from 32,768 (32k) to 1,024 (1k). After initial flagging for radio frequency interference (RFI), calibration solutions from primary and secondary calibrators were applied to the target field. The data were imaged using \textsc{Wsclean}\footnote{\url{https://gitlab.com/aroffringa/wsclean}} \citep{Offringa2014}, producing 10k × 10k pixel maps of the full bandwidth, as well as 8 sub-band images (68 MHz each) utilizing Briggs weighting, with (\textsc{Robust}) parameters of $-$0.5 and 0.0.

Direction independent self-calibration (2GC) was conducted with \textsc{Cubical}\footnote{\url{https://github.com/ratt-ru/CubiCal}} \citep{Kenyon2018} for delay and phase-only calibration, followed by amplitude and phase calibration using an iterative process of imaging and model prediction. The mask from the 1GC imaging was iteratively refined, for subsequent rounds of imaging.

For direction-dependent effect (DDE) corrections, we begin by `peeling' bright off-axis sources, which are numerous at UHF frequencies, coupled with the larger field of view (FoV). The peeling process involves phase-rotation and solving for gain terms using \textsc{Cubical}, with problematic sources iteratively uv-subtracted from the visibility data.

Finally, facet-based DDE corrections were applied using \textsc{DDFacet}\footnote{\url{https://github.com/saopicc/DDFacet}} \citep{Tasse2018} and \textsc{KillMS}\footnote{\url{https://github.com/saopicc/killMS}} \citep{Tasse2014, Smirnov2015}, dividing the sky model into multiple directions and solving for complex gains per facet. \textsc{DDFacet} was then used to apply the latter complex gain corrections on the fly, producing a direction-dependent calibrated (3GC) image with reduced artefacts around bright sources.

\section{Isolating Diffuse Radio Emission}
\label{sec3}

We identified the diffuse source through a systematic manual search for extended radio emission in the sample of 89 SPTpol galaxy clusters \citep{Huang_2020} that lie within the MeerKAT-SPT 100~deg$^2$ field. In this programme, each SZ-selected cluster was inspected for large-scale, low-surface-brightness radio features after compact sources were modeled and subtracted. The target discussed here revealed a clear excess of diffuse emission, consistent in extent and morphology with a radio halo.

To measure the integrated flux density of the diffuse emission, we must first subtract point sources in close proximity. This step, performed in the visibility domain, is essential to ensure that the flux density measurements of the diffuse emission were not contaminated by the embedded compact sources. Additionally, removing the point sources reduces blending with the extended emission when producing lower-resolution (high-$T_B$) maps. Point source subtraction was performed using \textsc{Crystalball}\footnote{\url{https://github.com/caracal-pipeline/crystalball}} package which utilizes a source list from \textsc{Wsclean} to generate visibilities containing only the identified point sources. The \textsc{Casa} \textsc{uvsub} task is then used to subtract these from the calibrated visbilities. We used these modified visibilities and modified imaging weights to produce lower-resolution high-$T_B$ maps that enhance the diffuse radio emission, as follows.

To locate diffuse radio emission in our target field, we take advantage of MeerKAT’s pinched core antenna configuration. At 816~MHz, short baselines within the dense core are sensitive to extended sources ($\gtrsim$ 1 arcmin), while longer baselines are used to identify and accurately subtract point sources. We impose a baseline-length lower limit ($uv$-range > 10 k$\lambda$) to carry out point-source substraction. The model from this image is used to create a point-source-subtracted image thereby isolating the diffuse emission.

The lower-resolution high-$T_B$ image was obtained by restricting the $uv$-range between 0-2~k$\lambda$ and using a Briggs \textsc{robust} = 0.0 in \textsc{DDFacet}, resulting in a beam size of approximately 12.7 arcsec $\times$ 9.93 arcsec.

\section{Results and Discussion}
\label{sec4}
\begin{figure*}
\begin{minipage}[!htb]{.54\textwidth}
\centering
\subfloat{\label{Fullband_img}\includegraphics[width=1\textwidth]{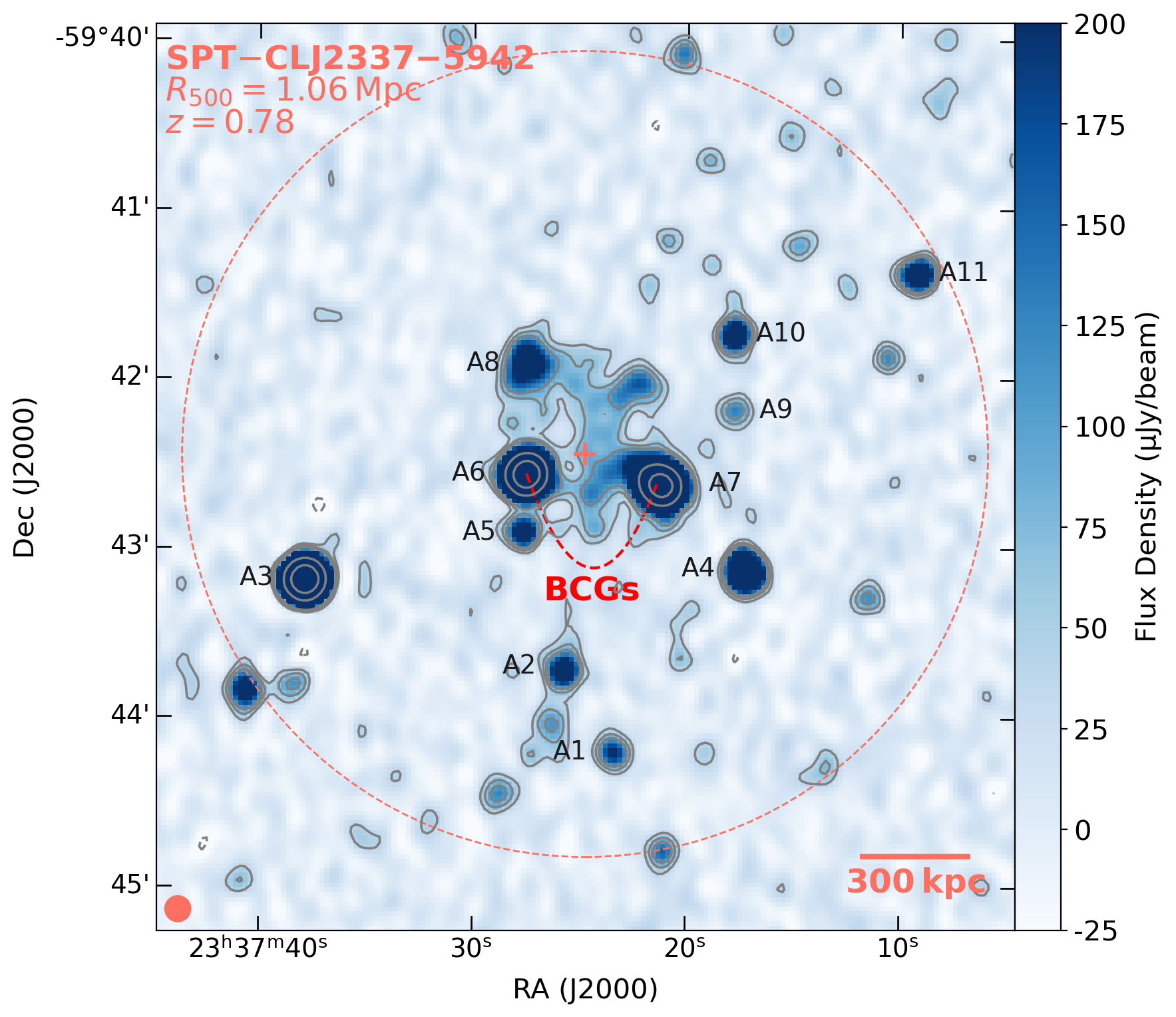}}
\end{minipage}%
\begin{minipage}[!htb]{.54\textwidth}
\centering
\subfloat{\label{Unifor_weighted}\includegraphics[width=1\textwidth]{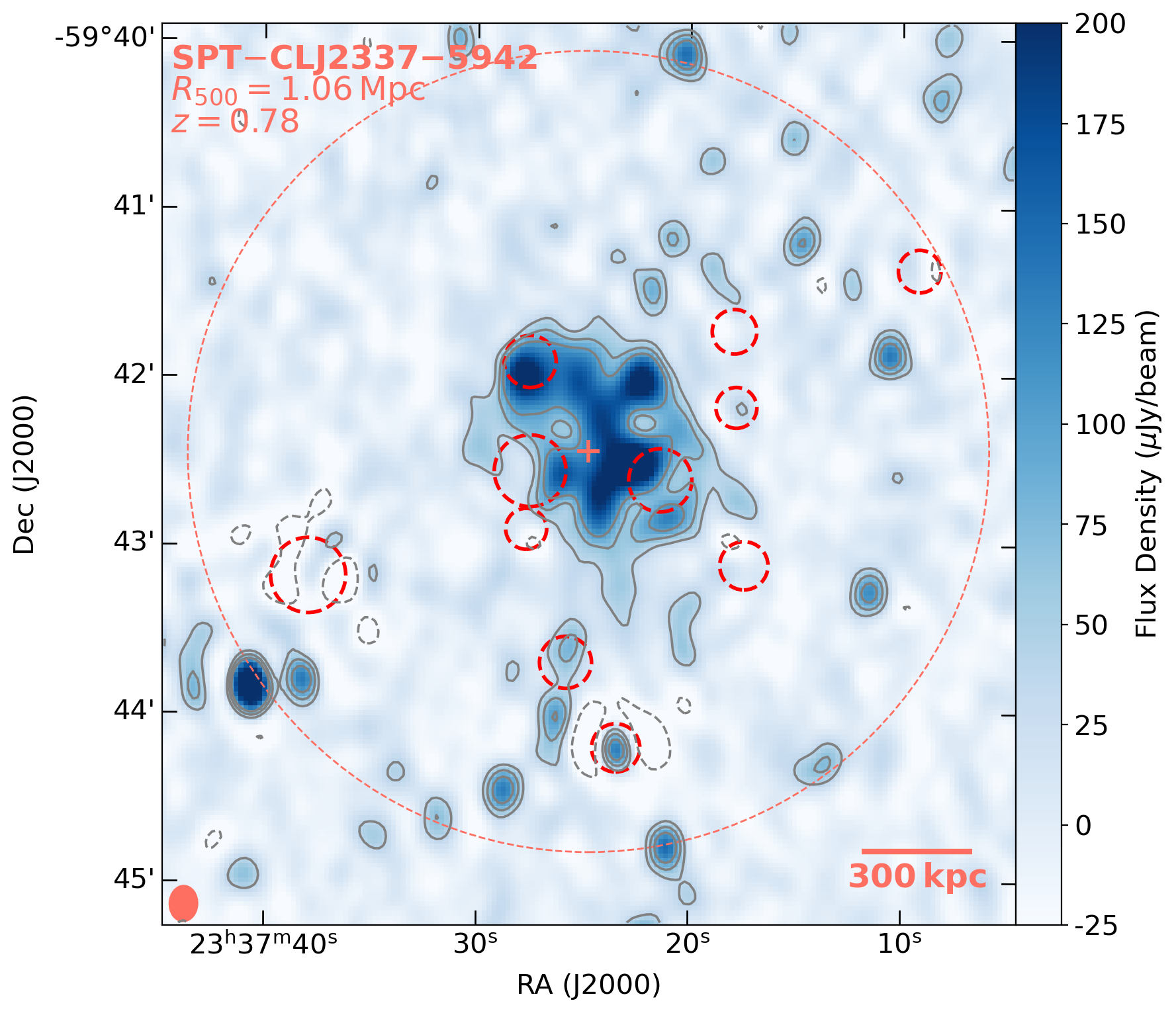}}
\end{minipage} 
\caption[UHF images]{Left: MeerKAT 816~MHz image of SPT-CLJ2337$-$5942 at \textsc{Robust} -0.5, with a resolution of 8.9 arcsec $\times$ 8.9 arcsec and position angle of 0.0 degree. Contour levels are drawn at [$-$5, 5, 9, 13, 17] $\times$ 1$\sigma$, where $\sigma$ = 9.15 $\mu$Jy beam$^{-1}$. The labels A1 - A11 (including the two BCGs) indicate the compact sources within the $R_{500}$ cluster region (orange circle). Right: Lower resolution (high-$T_B$) MeerKAT 816~MHz image of SPT-CLJ2337$-$5942 with discrete sources (denoted by small red circles) subtracted, at \textsc{Robust} 0.0, with a resolution of 12.6 arcsec $\times$ 9.9 arcsec and position angle of $-$2.17 degrees. Contour levels are drawn at [$-$5, 5, 9, 13, 17] $\times$ 1$\sigma$, where $\sigma$ = 7.50 $\mu$Jy beam$^{-1}$, with negative contours shown as dashed lines. The cluster center is marked by an orange `+'. The synthesized beams for both images are shown in orange in the lower-left corners.}
\label{HRandLR}
\end{figure*}

\begin{figure*}
\centering
\includegraphics[width=1.0\textwidth, trim=0 250 0 200,clip]{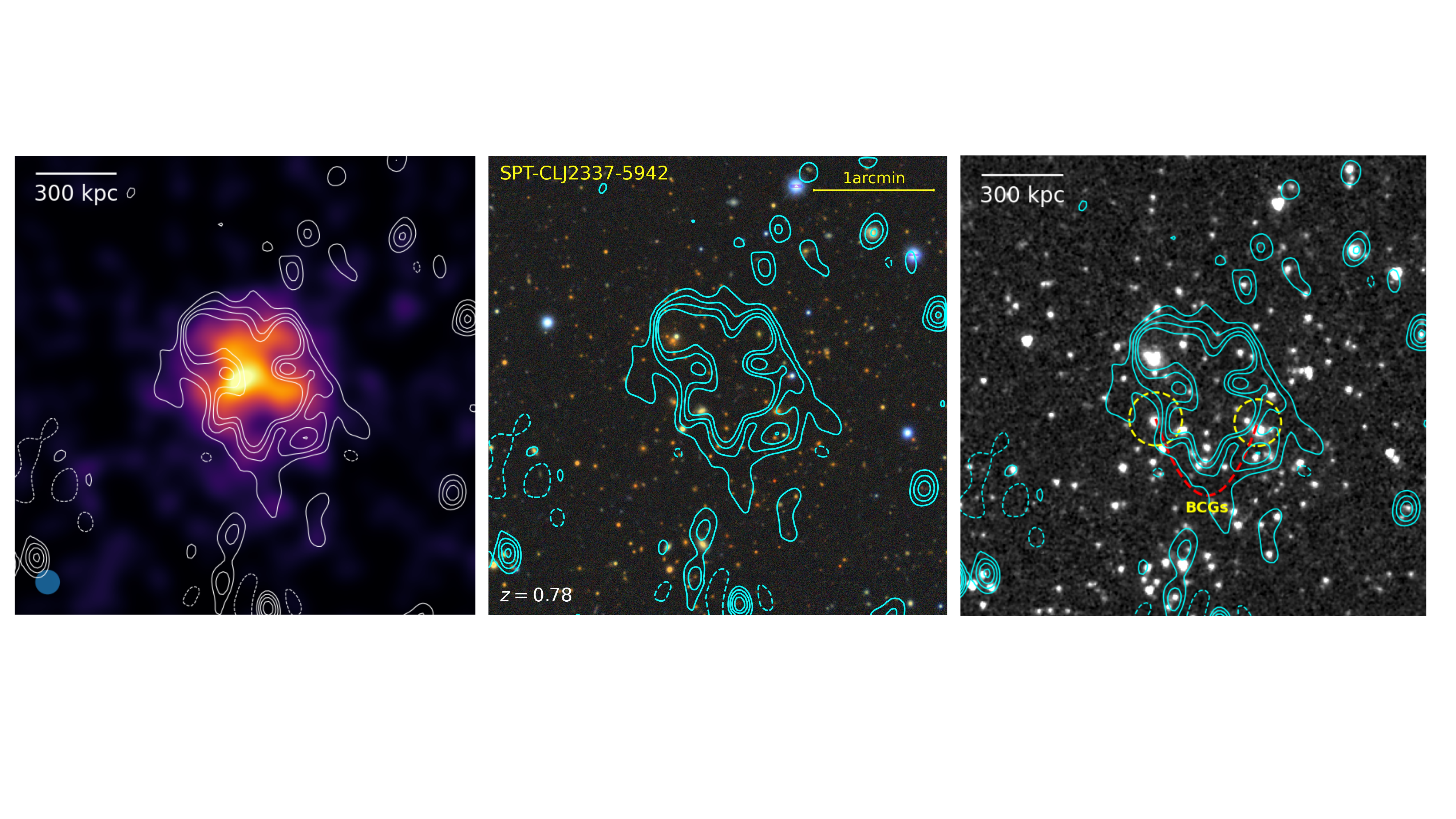}
\caption[Multiwave images]{Left: {\sl Chandra} X-ray image of SPT-CLJ2337$-$5942 in the 0.5 - 2.0 keV band, smoothed to 12 arcsec. The Gaussian smoothing kernel FWHM is indicated in the bottom-left corner. The radio halo emission is shown as white (teal) contours in the left (right) panel, with levels at [$-$5, 5, 9, 13, 17] $\times$ 1$\sigma$, where $\sigma$ = 7.50 $\mu$Jy beam$^{-1}$. The emission traces the brightness peak and exhibits a filamentary morphology. Middle: Optical grz image of SPT-CLJ2337$-$5942 from the DESI-Legacy-Surveys. The radio halo overlaps with two bright galaxies which are possible BCGs (e.g., \citealt{Andersson_2011}) counterparts embedded in the halo. Right: Spitzer/IRAC 3.6 $\mu$m image \citep{Ashby_2013} with MeerKAT contours, highlighting the overdensity of mid-infrared (MIR)-detected cluster galaxies. The IRAC panel also shows that several compact radio peaks have IR counterparts, while the diffuse radio halo emission extends over the same region as the cluster galaxy distribution.} 
\label{Multiwave}
\end{figure*}

The MeerKAT UHF-band high-$T_B$ image reveals extended diffuse emission detected with high confidence at the centre of SPT-CLJ2337$-$5942. The largest linear size (LLS) is $\sim$800~kpc at 816~MHz and has a complex morphology (see Figure \ref{HRandLR}). We compare this radio image with a {\sl Chandra} X-ray 0.5 - 2 keV image, obtained from archival {\sl Chandra} ACIS-I data (ObsID 11859; PI: Garmire) with a total exposure time of 19.77 ksec. As seen in Figure~\ref{Multiwave}, the diffuse radio emission has a very similar morphological structure to the {\sl Chandra} X-ray emission, presumed to trace the ICM. Based on the cluster location, the close correspondence between diffuse X-ray and radio emission, we classify this diffuse emission as a radio halo. The basic properties of the cluster are presented in Table \ref{properties}.\\\\

\begin{table*}
\caption{The properties of the compact sources within the $R_{500}$ cluster region. The BCGs are denoted by $\textbf{*}$.}
\small 
\centering
\begin{tabular}{cccccc}
\toprule\toprule
\textbf{ID} & \textbf{RA (J2000)} & \textbf{Dec (J2000)} & \textbf{$z_{\text{phot}}$} & \textbf{Flux} & \textbf{Notes} \\
\midrule
& \text{(hms)} & \text{(dms)} & & \text{(mJy)} &  \\
\midrule
A1 & 23 37 17 & $-$59 43 08 & 0.76 & 0.83 $\pm$ 0.02 & Cluster member \\
A2 & 23 37 38 & $-$59 43 11 & 0.78 & 6.7 $\pm$ 0.03 & Cluster member \\
A3 & 23 37 26 & $-$59 43 43 & 0.27 & 0.36 $\pm$ 0.02 & Foreground galaxy \\
A4 & 23 37 18 & $-$59 42 12 & 0.76 & 0.12 $\pm$ 0.02 & Cluster member \\
A5 & 23 37 23 & $-$59 44 13 & 1.2 & 0.22 $\pm$ 0.02 & Background galaxy \\
A6 & 23 37 28 & $-$59 42 33 & 0.78 & 6.6 $\pm$ 0.03 & Cluster member $\textbf{*}$ \\
A7 & 23 37 21 & $-$59 42 37 & 0.81 & 5.8 $\pm$ 0.03 & Cluster member $\textbf{*}$ \\
A8 & 23 37 27 & $-$59 41 54 & 0.97 & 0.57 $\pm$ 0.02 & Background galaxy \\
A9 & 23 37 16 & $-$59 42 10 & 0.84 & 0.26 $\pm$ 0.02 & Background galaxy \\
A10 & 23 37 15 & $-$59 41 43 & 1.0 & 0.34 $\pm$ 0.02 & Background galaxy \\
A11 & 23 37 07 & $-$59 41 22 & 1.0 & 0.34 $\pm$ 0.02 & Background galaxy \\
\bottomrule
\end{tabular}
\label{CompactSources}
\end{table*}

An optical RGB image ({\sl grz}-band) from DESI Legacy Survey \citep{Dey_2019}, shown in Figure \ref{Multiwave}, shows that SPT-CLJ2337$-$5942 hosts two brightest cluster galaxies (BCGs) and multiple sub-groups undergoing collision with the radio halo, further reinforcing its classification as a merging cluster (disturbed). \citet{Yuan_2020} and \citet{Yuan_2022} confirmed that SPT-CLJ2337$-$5942 is a disturbed galaxy cluster based on its X-ray morphology. The optical and infrared (IR) images (see Figure \ref{Multiwave}) further highlight the galaxy distribution, which appears elongated along the merger axis and consistent with the disturbed morphology of the system. Within the $R_{500}$ = 1.1 Mpc radius, we identify eleven compact radio sources in our 816~MHz full-resolution image (labelled A1 - A11 in Figure \ref{HRandLR}). Of these, five are confirmed as cluster members (including the two BCGs), five are associated with background galaxies ($z> 0.8$), and one is a foreground galaxy ($z\sim 0.272$), based on photometric redshifts from the DESI Legacy Imaging Surveys Data (DR8; \citealt{Duncan_2022}). The properties of these compact sources are detailed in Table \ref{CompactSources}.\\\\
The halo shows an overall smooth, diffuse morphology, but with noticeable filamentary substructures that trace the thermal X-ray emission of the ICM, as shown in Figure \ref{Multiwave}. The striking spatial correlation between the diffuse synchrotron emission and the X-ray brightness strongly supports a scenario in which both components arise from the same large-scale turbulent environment. In this framework, ICM turbulence-driven by cluster mergers both heats the ICM gas and re-accelerates cosmic-ray electrons, with magnetic fields permeating the ICM enabling synchrotron emission across the X-ray-emitting volume \citep[e.g.,][]{Brunetti_2014}. This interpretation is supported by quantitative studies that report point-to-point correlations between radio and X-ray surface brightness in several clusters hosting radio halos \citep[e.g.,][]{Govoni_2001,Feretti_2001,Giacintucci_2005,Brown_2011coma,Rajpurohit_2018} and it is self-evident that the point-to-point correlation in SPT-CLJ2337$-$5942 is very high. However, exceptions exist where no such correlation is found \citep[e.g.,][]{Shimwell_2014}, highlighting the complexity of radio-X-ray connections in the ICM and the importance of deep, high-fidelity radio imaging.

We quantified the radio-X-ray connection by performing a pixel-by-pixel correlation analysis between the reprojected maps (Figure not shown). We find a strong Pearson correlation coefficient of $r=0.73$ ($\rho$-value $\ll10^{-10}$) and a Spearman rank correlation coefficient of $\rho=0.50$, both indicating a highly significant correlation between the diffuse radio and thermal X-ray surface brightness. Fitting a power-law of the form $I_\mathrm{radio} \propto I^{b}_{X}$, we obtain a best-fit slope of $b=0.72\pm0.02$. This slope is fully consistent with point-to-point correlations measured in other clusters hosting radio halos \citep[e.g.,][]{Govoni_2001,Feretti_2001,Giacintucci_2005,Brown_2011coma,Rajpurohit_2018}, further reinforcing the picture that the radio and X-ray emission in SPT-CLJ2337$-$5942 trace the same turbulent ICM volume.

\begin{figure}
\centering
\includegraphics[width=0.98\linewidth]{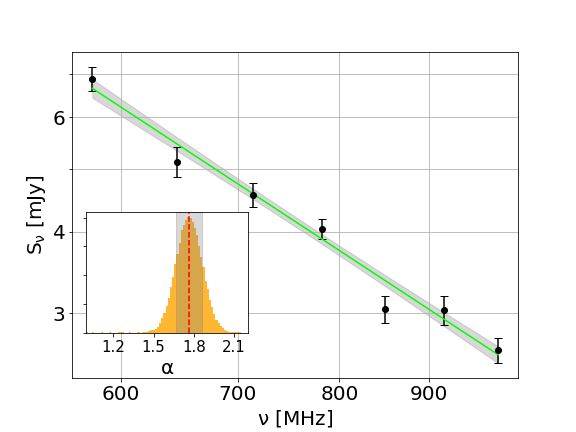}
\caption[Spectral index plot]{Integrated radio spectrum of the halo between $\nu = 578$-$986$~MHz. The points show the measured flux densities with $1\sigma$ uncertainties. The green line indicates the median posterior-fit power-law derived from an MCMC sampler \citep{emcee}, with the shaded region representing the 68 percent confidence intervals. The integrated spectral index is $\alpha = 1.76 \pm 0.10$, classifying the source as an ultra-steep spectrum radio halo (USSRH). The inset in the bottom-left corner shows the posterior probability distribution of $\alpha$ derived from the MCMC sampling.}
\label{SPI}
\end{figure}

Additionally, there appear to be low-brightness depressions in the X-ray emission that resemble cavities. These are unlikely to be AGN-inflated bubbles, given the absence of appropriately located compact radio cores, jets, or optical AGN features. Alternatively, these structures can  be interpreted as merger-induced low-density regions shaped by bulk ICM motions or turbulence \citep[e.g.,][]{Markevitch_2007,Vazza_2009,ZuHone_2011}. The latter is consistent with the radio halo scenario, in which the synchrotron-emitting electrons are re-accelerated over large volumes without requiring ongoing AGN activity, particularly in post-merger or dynamically evolved systems \citep[e.g.,][]{Brunetti_2008}.\\\\
The flux density of the radio halo is measured as $S_{816}$ = 4.34 $\pm$ 0.41 mJy. From sub-band image flux density measurements, we calculate the integrated spectral index of the radio halo to be $\alpha$ $\sim$ 1.76 $\pm$ 0.10 (between $\nu =  578 - 986$~MHz) using a Markov Chain Monte Carlo (MCMC) sampler \citep{emcee}. The integrated spectral index posterior probability distribution function (PDF, see Figure \ref{SPI}) identifies the halo as an ultra-steep spectrum radio halo (USSRH) with high confidence. As discussed in Sec.~\ref{sec:intro}, this is a rare type of halo, particularly at $z\gtrsim 0.5$. USSRHs are thought to be non-thermal emission generated by aged electron populations following rare minor cluster mergers in massive systems (see \citealt{Cassano_2008b,Cassano_2009,Cassano_2010b} for more details). These halos typically exhibit low surface brightness (e.g., \citealt{Weeren_2019}), consistent with SPT-CLJ2337$-$5942, for which we estimate a surface brightness of $\sim$0.4 $\mu$Jy arcsec$^{-2}$, which is in line with expectations (e.g., \citealt{pandey_2016,Rajpurohit_2022,santra2023}). This low surface brightness, combined with their steep spectra, often causes USSRHs to be missed in wide-area surveys, as they require highly sensitive, high-fidelity images at frequencies $v \lesssim$ 1 GHz. SPT-CLJ2337$-$5942 is the most distant USSRH known to date, at a redshift of $z$ = 0.78. This detection therefore demonstrates that the sensitivity and image fidelity of the MeerKAT UHF band can reveal distant USSRHs in relatively short observations ($\sim$1 hr), enabling large-area surveys conducive to the discovery of rare systems such as this. At the same time, the discovery within a 100~deg$^2$ area suggests that USSRHs could be more common than the current total sample ($\sim$~20) indicates, though this may depend on observational sensitivity and selection effects (e.g., \citealt{Cassano_2012,Pasini_2024}).\\\\

Using the integrated flux density and spectral index of the halo, we estimate a radio power of approximately $P_{1.4} = 9.6 \pm 0.9 \times 10^{23}$~W\,Hz$^{-1}$, scaled from the observed frequency of 816~MHz assuming the in-band spectral index. At high redshifts, relativistic electrons experience energy losses due to inverse Compton (IC) scattering with cosmic microwave background (CMB) photons, which are expected to shorten their radiative lifetimes and reduce the detectability of synchrotron-emitting regions (e.g., \citealt{Ferrari_2008,Cassano_2009, 2014_Lindner}). The low radio power may also be due to the dominance of IC losses at high redshift, which arise from the higher density and temperature of CMB photons at earlier cosmological times (e.g., \citealt{Sweijen_2022}). The flux and its uncertainties are calculated using the \textsc{Radioflux}\footnote{\url{https://github.com/mhardcastle/radioflux}} radio flux measurement tool.

\begin{figure}
\centering
\includegraphics[width=1\linewidth, trim=450 0 450 0,clip]{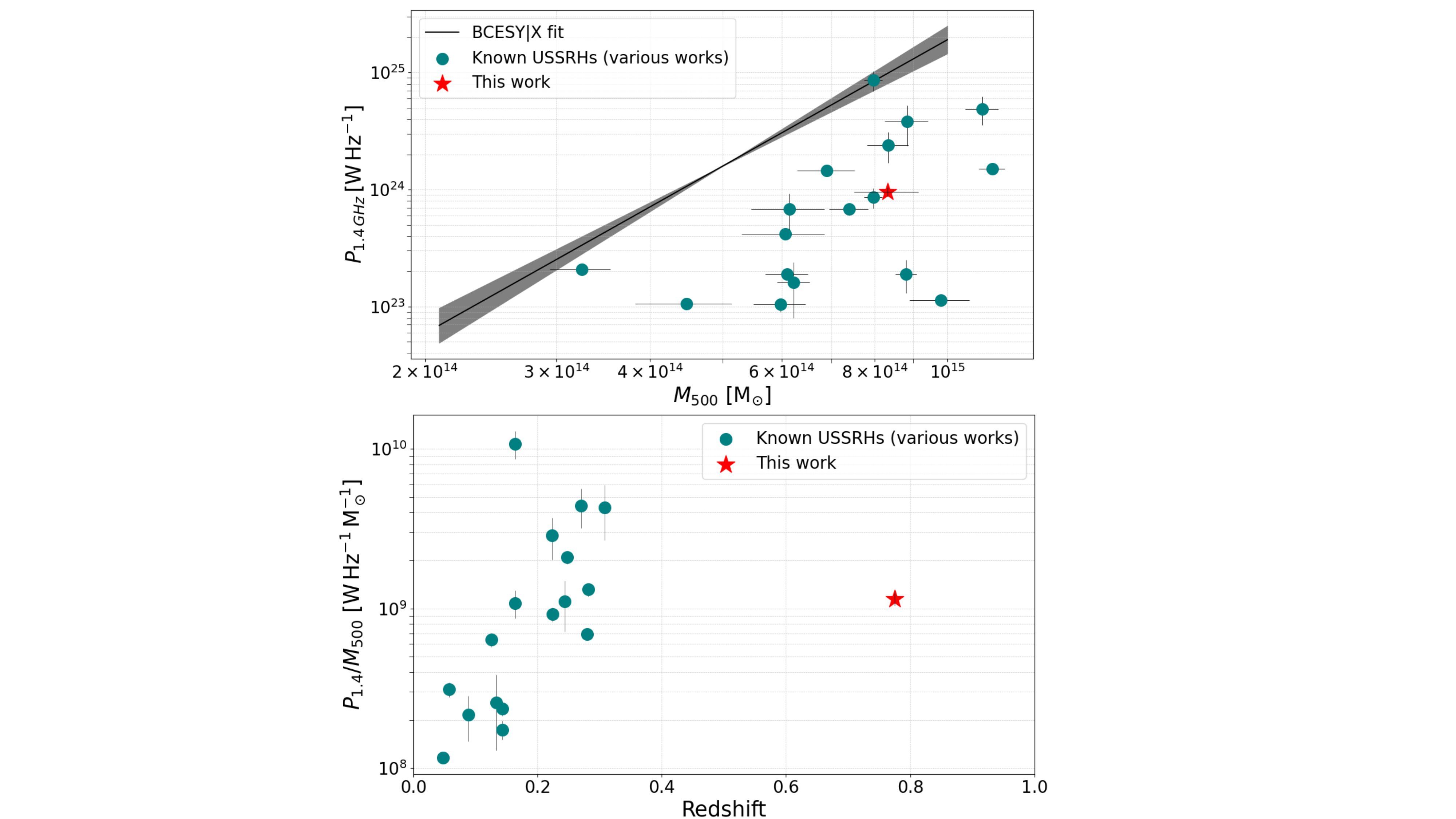}
\caption[Scaling relation]{Top: The $P_{1.4\,\mathrm{GHz}}$ - $M_{500}$ scaling relation of known USSRHs \citep{Cuciti_2021,Duchesne2021,Rajpurohit_2022,Pasini_2022,Edler_2022,Venturi_2022}. The USSRH presented in this work is marked with a red star. The solid line represents the correlation for normal and ultra-steep spectrum radio halos, as reported by \citet{Cuciti_2021b}. Bottom: Radio power at 1.4 GHz normalized by cluster mass, $P_{1.4}/M_{500}$, for known USSRHs as a function of redshift.}
\label{ScalingRelation}
\end{figure}

We compare this radio power to the sample of known USSRHs ($z$ $\sim$ 0.048 - 0.4) in the literature. The $P_{1.4}$ - $M_{500}$ scaling relation (see Figure \ref{ScalingRelation}) shows that the USSRH radio power lies within the scatter of both nearby and distant USSRH populations, occupying a region below the ordinary halo correlation. By normalizing halo power by cluster mass, we can test whether our system is unusual compared to known USSRHs. Figure \ref{ScalingRelation} further illustrates that while most USSRHs are found at $z \lesssim 0.4$, our detection at $z \sim 0.78$ extends this population to much higher redshift. Its $P_{1.4}/M_{500}$ ratio is consistent with lower-z USSRHs, indicating that the underlying particle acceleration mechanisms remain comparably efficient even in the early Universe.

Assuming turbulent re-acceleration remains the dominant mechanism \citep{Brunetti_2000,Cassano_2006,Brunetti_2008,Cassano_2009,Brunetti_2014}, such a trend would suggest that stronger magnetic fields and enhanced turbulence are required at higher redshifts \citep{Di_Gennaro_2025}. This is particularly relevant given that inverse-Compton losses increase with redshift (e.g., \citealt{Ferrari_2008,Cassano_2009, 2014_Lindner}), reducing the efficiency of synchrotron emission. Thus, sustaining radio halo luminosities comparable to low-redshift systems \citep{Cuciti_2021,Duchesne2021,Rajpurohit_2022,Pasini_2022,Edler_2022,Venturi_2022} would necessitate a greater injection of turbulent energy or elevated magnetic field strengths in high-$z$ clusters, as anticipated by turbulent re-acceleration models \citep[e.g.,][]{Gennaro2021}. This is shown in Figure~\ref{ScalingRelation}, where the best-fit line is based on the scaling relation studies of USSRHs conducted by \citet{Cuciti_2021b}, following the fitting procedure outlined in \citet{Cassano_2013}. The scaling relation further supports the USSRH classification, as the halo appears radio under-luminous in comparison to the correlation reported by \citet{Cuciti_2021b} between 1.4 GHz radio power and the cluster mass.
The turbulent re-acceleration model uniquely predicts much steeper spectra, as a result of less energetic cluster merger events (e.g.,  \citealt{Cassano_2006,Brunetti_2008,Cassano_2009}). Therefore, the detection of the USSRH is significant evidence supporting this model over the secondary electron origin (hadronic model; \citealt{Dennison_1980,Blasi_1999}), which requires a substantial proton energy budget (e.g., \citealt{Brunetti_2004}). In merging clusters, steep-spectrum halos are predicted to result from turbulence-driven acceleration of relativistic electrons \citep{Brunetti_2000,Brunetti_2014}, and ongoing/forthcoming low-frequency arrays, together with wide-field, high-fidelity surveys like MeerKAT-SPT, will enable observational tests of these theoretical predictions \citep[e.g.,][]{Macario_2010,Cassano_2012,Pasini_2024}.

\section{Conclusion}
\label{sec5}
In this Letter, we present MeerKAT UHF-band observations of the galaxy cluster SPT-CLJ2337$-$5942 at {$z=0.78$}. The high-$T_B$ images reveal an USSRH within the cluster, marking the highest redshift for an ultra-steep spectrum radio halo (USSRH) reported to date. These are important systems that enable us to discern between radio halo formation models, and SPT-CLJ2337$-$5942 has two key traits in this regard:
\begin{enumerate}
    \item An integrated spectral index of $\alpha \sim 1.76 \pm 0.10$ (578-986 MHz), significantly steeper than that predicted by hadronic models, indicating that ongoing particle re-acceleration is necessary.
    \item A striking spatial correlation between the filamentary-like diffuse radio emission and the X-ray brightness, implying a direct link between thermal ICM electrons, non-thermal electrons, and cluster magnetic fields.
\end{enumerate}
This USSRH has a radio power of $P_{1.4} = 9.6 \pm 0.9 \times 10^{23}$~W\,Hz$^{-1}$. According to the $P_{1.4}$ - $M_{500}$ scaling relation, the USSRH appears under-luminous, as it may result from less energetic phenomena, such as minor cluster mergers within more massive systems (e.g., \citealt{Cassano_2008b,Cassano_2010b}). These processes typically lead to less luminous steep spectrum halos ($\alpha$ > 1.5). Given the steep spectrum of the USSRH in SPT-CLJ2337$-$5942, a hadronic origin for the halo is unlikely. Instead, turbulence re-acceleration is likely the primary mechanism for halo production, with the necessary energy for this re-acceleration expected to be supplied by cluster merger events. Future observations with the SKA, together with ongoing low-frequency surveys such as MeerKAT-SPT, will significantly expand the sample of high-redshift USSRHs, enabling more robust tests of radio halo formation models and particle acceleration mechanisms.

\section*{Acknowledgements}

We acknowledge the financial support of the South African Radio Astronomy Observatory (SARAO) for this research. The MeerKAT telescope is operated by SARAO, a facility of the National Research Foundation (NRF), which is an agency under the Department of Science and Innovation (DSI). We also express our gratitude to the SARAO science commissioning and operations team, led by Sharmila Goedhart, for their assistance. We also acknowledge the use of the ilifu cloud computing facility (\url{www.ilifu.ac.za}), a partnership between the University of Cape Town, the University of the Western Cape, Stellenbosch University, Sol Plaatje University, and the Cape Peninsula University of Technology. The ilifu facility is supported by contributions from the Inter-University Institute for Data Intensive Astronomy (IDIA), a partnership between the University of Cape Town, the University of Pretoria, and the University of the Western Cape; the Computational Biology division at UCT; and the Data Intensive Research Initiative of South Africa (DIRISA). MA is supported by FONDECYT grant number 1252054, and gratefully acknowledges support from ANID Basal Project FB210003 and ANID MILENIO NCN2024$\_$112.
\\\\
Additionally, this work made use of the CARTA (Cube Analysis and Rendering Tool for Astronomy) software (DOI: \url{10.5281/zenodo.3377984} - \url{https://cartavis.github.io}).

\section*{Data Availability}

The MeerKAT raw data used in this study are publicly available (Project IDs SCI-20220822-JV-02 and SCI-20230907-JV-01) in accordance with the South African Radio Astronomy Observatory data release policy, while calibrated and advanced data products are available upon reasonable request.


\bibliographystyle{mnras}
\bibliography{example}







\bsp	
\label{lastpage}
\end{document}